\documentstyle[11pt,psfig]{article}

\def\Re{{\cal R \mskip-4mu \lower.1ex \hbox{\it e}\,}}
\def\Im{{\cal I \mskip-5mu \lower.1ex \hbox{\it m}\,}}
\def\ie{{\it i.e.}}
\def\eg{{\it e.g.}}
\def\etc{{\it etc}}

\def\sub#1{_{\lower.25ex\hbox{$\scriptstyle#1$}}}
\def\tev{\,{\ifmmode\mathrm {TeV}\else TeV\fi}}
\def\gev{\,{\ifmmode\mathrm {GeV}\else GeV\fi}}
\def\mev{\,{\ifmmode\mathrm {MeV}\else MeV\fi}}
\def\mpl{\ifmmode M_{pl}\else $M_{pl}$\fi}
\def\mpl{\ifmmode \overline M_{Pl}\else $\bar M_{Pl}$\fi}
\def\to{\rightarrow}

\def\subw{_{\rm w}}
\def\mh{\ifmmode m\sbl H \else $m\sbl H$\fi}
\def\mch{\ifmmode m_{H^\pm} \else $m_{H^\pm}$\fi}
\def\mt{\ifmmode m_t\else $m_t$\fi}
\def\mc{\ifmmode m_c\else $m_c$\fi}
\def\mz{\ifmmode M_Z\else $M_Z$\fi}
\def\mw{\ifmmode M_W\else $M_W$\fi}
\def\mws{\ifmmode M_W^2 \else $M_W^2$\fi}
\def\mhs{\ifmmode m_H^2 \else $m_H^2$\fi}   
\def\mzs{\ifmmode M_Z^2 \else $M_Z^2$\fi}
\def\mts{\ifmmode m_t^2 \else $m_t^2$\fi}
\def\mcs{\ifmmode m_c^2 \else $m_c^2$\fi}
\def\mchs{\ifmmode m_{H^\pm}^2 \else $m_{H^\pm}^2$\fi}
\def\ztwo{\ifmmode Z_2\else $Z_2$\fi}
\def\zone{\ifmmode Z_1\else $Z_1$\fi}
\def\mtwo{\ifmmode M_2\else $M_2$\fi}
\def\mone{\ifmmode M_1\else $M_1$\fi}
\def\tb{\ifmmode \tan\beta \else $\tan\beta$\fi}
\def\xw{\ifmmode x\subw\else $x\subw$\fi}
\def\ch{\ifmmode H^\pm \else $H^\pm$\fi}
\def\lum{\ifmmode {\cal L}\else ${\cal L}$\fi}
\def\inpb{\,{\ifmmode {\mathrm {pb}}^{-1}\else ${\mathrm {pb}}^{-1}$\fi}}
\def\infb{\,{\ifmmode {\mathrm {fb}}^{-1}\else ${\mathrm {fb}}^{-1}$\fi}}
\def\epem{\ifmmode e^+e^-\else $e^+e^-$\fi}
\def\ppb{\ifmmode \bar pp\else $\bar pp$\fi}
\def\bsg{\ifmmode B\to X_s\gamma\else $B\to X_s\gamma$\fi}
\def\bsll{\ifmmode B\to X_s\ell^+\ell^-\else $B\to X_s\ell^+\ell^-$\fi}
\def\bstt{\ifmmode B\to X_s\tau^+\tau^-\else $B\to X_s\tau^+\tau^-$\fi}
\def\lamt{\ifmmode \tilde\lambda\else $\tilde\lambda$\fi}
\def\shat{\ifmmode \hat s\else $\hat s$\fi}
\def\that{\ifmmode \hat t\else $\hat t$\fi}
\def\uhat{\ifmmode \hat u\else $\hat u$\fi}

\newskip\zatskip \zatskip=0pt plus0pt minus0pt
\def\matth{\mathsurround=0pt}
\def\lsim{\mathrel{\mathpalette\atversim<}}

\def\atversim#1#2{\lower0.7ex\vbox{\baselineskip\zatskip\lineskip\zatskip
  \lineskiplimit 0pt\ialign{$\matth#1\hfil##\hfil$\crcr#2\crcr\sim\crcr}}}

\def\grtsim{\,\,\rlap{\raise 3pt\hbox{$>$}}{\lower 3pt\hbox{$\sim$}}\,\,}
\def\lsim{\,\,\rlap{\raise 3pt\hbox{$<$}}{\lower 3pt\hbox{$\sim$}}\,\,}

\renewcommand{\thefootnote}{\fnsymbol{footnote}}

\begin{document} \begin{titlepage}
\rightline{\vbox{\halign{&#\hfil\cr
&SLAC-PUB-11930\cr
}}}
\begin{center}
\thispagestyle{empty} \flushbottom { {
\Large\bf Higher Curvature Effects in ADD and RS Models
\footnote{Work supported in part
by the Department of Energy, Contract DE-AC02-76SF00515}
\footnote{e-mail:
$^a$rizzo@slac.stanford.edu}}}
\medskip
\end{center}

\centerline{Thomas G. Rizzo$^{a}$}
\vspace{8pt} 
\centerline{\it Stanford Linear
Accelerator Center, 2575 Sand Hill Rd., Menlo Park, CA, 94025}

\vspace*{0.3cm}

\begin{abstract}
Over the last few years several extra-dimensional models have been introduced 
in attempt to deal with the hierarchy problem. These models can lead to 
rather unique and spectacular signatures at Terascale colliders such as the 
LHC and ILC. The ADD and RS models, though quite distinct, have many common 
feature including a constant curvature bulk, localized Standard Model(SM) 
fields and the assumption of the validity of the EH action as a description 
of gravitational interactions. 
\end{abstract}

\vspace*{3.0in}
\noindent
Talk given at the {\it International Linear Collider Workshop-LCWS06}, 
Bangalore, India, 16-18 March, 2006.


\renewcommand{\thefootnote}{\arabic{footnote}} \end{titlepage} 

%
%
%

\section{Introduction}

Over the last few years several extra-dimensional models have been introduced 
in attempt to deal with the hierarchy problem. These models can lead to 
rather unique and spectacular signatures at Terascale colliders such as the 
LHC and ILC. The ADD and RS models, though quite distinct, have many common 
feature including a constant curvature bulk, localized Standard Model(SM) 
fields and the assumption of the validity of the EH action as a description 
of gravitational interactions. 

It is well known that the EH action can only be regarded as an effective 
theory so we may on general grounds  
expect that as one probes energies approaching the fundamental scale, 
$M$, (as one will do at the LHC/ILC for both models) significant deviations 
from EH expectations are likely to appear. From a bottom-up approach we can 
claim ignorance of what a more complete gravitational theory may be like but 
we would expect that the leading 
deviations from EH would appear as higher dimensional 
operators involving various invariants constructed from the curvature tensor.  
This motivates us to consider a more general class of gravitational actions 
of the $D=4+n$ dimensional form 
\begin{equation}
S_g={{M^{D-2}}\over {2}} \int d^D x {\sqrt g} ~F(R,P,Q)\,,
\end{equation}
where $P=R_{AB}R^{AB}$ and $Q=R_{ABCD}R^{ABCD}$. Various actions of this 
generic type have been examined in the literature within a variety of 
contexts including black holes, cosmology, modifications of gravity 
within the solar system as well as at short-distances scales of interest 
to us here{\cite {me}}. It will be assumed in what follows that at low 
energies $F\to R$ so that the EH action, with all its successes, is recovered. 

In both the ADD and RS models we wish 
to know the graviton KK spectrum and wavefunctions as well as their couplings 
to SM fields. We also need to know how \mpl and the more fundamental scale 
$M$ are related. To this end it can be shown that for spaces with constant 
curvature the above action is the most general one; furthermore it can 
be shown that we can obtain all of the desired information by expanding 
this action around 
the constant background metric to quadratic order{\cite {me}}. Performing 
this expansion then yields an effective action of the form 
\begin{equation}
S_{eff}={{M^{D-2}}\over {2}} \int d^D x {\sqrt g} ~\big[\Lambda+a_1R+
a_2R^2+a_3C+a_4G\big]\,,
\end{equation}
where 
$G=R^2-4R_{AB}R^{AB}+R_{ABCD}R^{ABCD}=R^2-4P+Q$ is the Gauss-Bonnet invariant 
and 
$C=C_{ABCD}C^{ABCD}=Q-4P/(n+2)+2R^2/(n+3)(n+2)$
is the square of the Weyl tensor. This simplification is obtainable from the 
direct expansion $F=F_0+(R-R_0)F_R+(P-P_0)F_P+(Q-Q_0)F_Q+{\rm {quadratic~ 
terms}}$, where $F_0$ is a constant corresponding to the evaluation of 
$F$ itself in the fixed curvature background metric and 
$F_X=\partial F/\partial X|_0$; a quantity  $X_0$ here means that $X$ 
is to be evaluated in terms of the background metric 
which we here assume to be a space of constant curvature as is the case in 
both the ADD and RS models. Thus the quantities $R_0$, $P_0$, $Q_0$, $F_X$ 
and $F_{XY}$ are just numbers which depend on the explicit form of the 
metric and the number of extra dimensions. This procedure directly gives us 
the constants $\Lambda,a_i$ above{\cite {me}} in terms of $F$ and its 
derivatives evaluated in the background metric as we will see below.

\section{ADD}

For the action $F$, the D-dimensional field content consists of a massless 
tensor field (\ie, the `usual' graviton) as well as a massive scalar field and 
a massive tensor 
ghost field with bulk masses directly calculable from $F$ itself. This can be 
seen most clearly in, \eg, the ADD model where the full exchange amplitude 
between two localized SM sources (before KK summation) can be written as 
$A \sim [T_{\mu\nu}T^{\mu\nu}-T^2/(n+2)]/(k^2-m_n^2)$$-[T_{\mu\nu}T^{\mu\nu}-
T^2/(n+3)]/(k^2-(m_T^2+m_n^2))$$+T^2/(n+2)(n+3)[k^2-(m_S^2+m_n^2)]$. 
Here $T_{\mu\nu}$ is the localized SM stress-energy tensor and $T$ is its 
trace. Ghost fields are potentially very dangerous in perturbation theory and 
can cause unitarity violations so we may want to rid ourselves of this 
massive tensor field. 
To remove this ghost state from the spectrum it is sufficient to force the 
bulk tensor mass $m_T$ to infinity; this will make all of the fields in 
its entire KK tower decomposition have infinite mass. In this simple ADD 
case one finds that 
$m_T^2=-F_R/(n+2)(F_P+4F_Q)$ and thus we see that to make $m_T$ infinitely 
large it is sufficient to demand that $F$ be only a function of $R$ and the 
combination $Q-4P$. These results can also be shown to hold qualitatively 
in the case of the RS model and the $Q-4P$ dependence of $F$ 
will be assumed in the discussion below. Of course, we still need to insure 
that the usual KK gravitons and the new KK scalars are non-tachyonic so that 
$F_R>0$ \etc will also be required. For either ADD or RS model, one now in 
general finds that $a_3=0$ and 
\begin{eqnarray}
\Lambda&=&F_0-R_0F_R+F_{RR}R_0^2/2+\lambda R_0^2[F_Q-R_0F_{QQ}+
\lambda F_{QQ}R_0^2/2]\nonumber \\
a_1&=&F_R-R_0F_{RR}+\lambda R_0^2F_{RQ}\nonumber \\
a_2&=&-F_Q+F_{RR}/2+R_0F_{RQ}-\lambda R_0^2F_{QQ}\nonumber \\
a_4&=&-a_2+F_{RR}/2\,.
\end{eqnarray}
\begin{figure}[htbp]
\centerline{
\psfig{figure=shift.ps2,height=7.5cm,width=7.5cm,angle=90}
\hspace*{5mm}
\psfig{figure=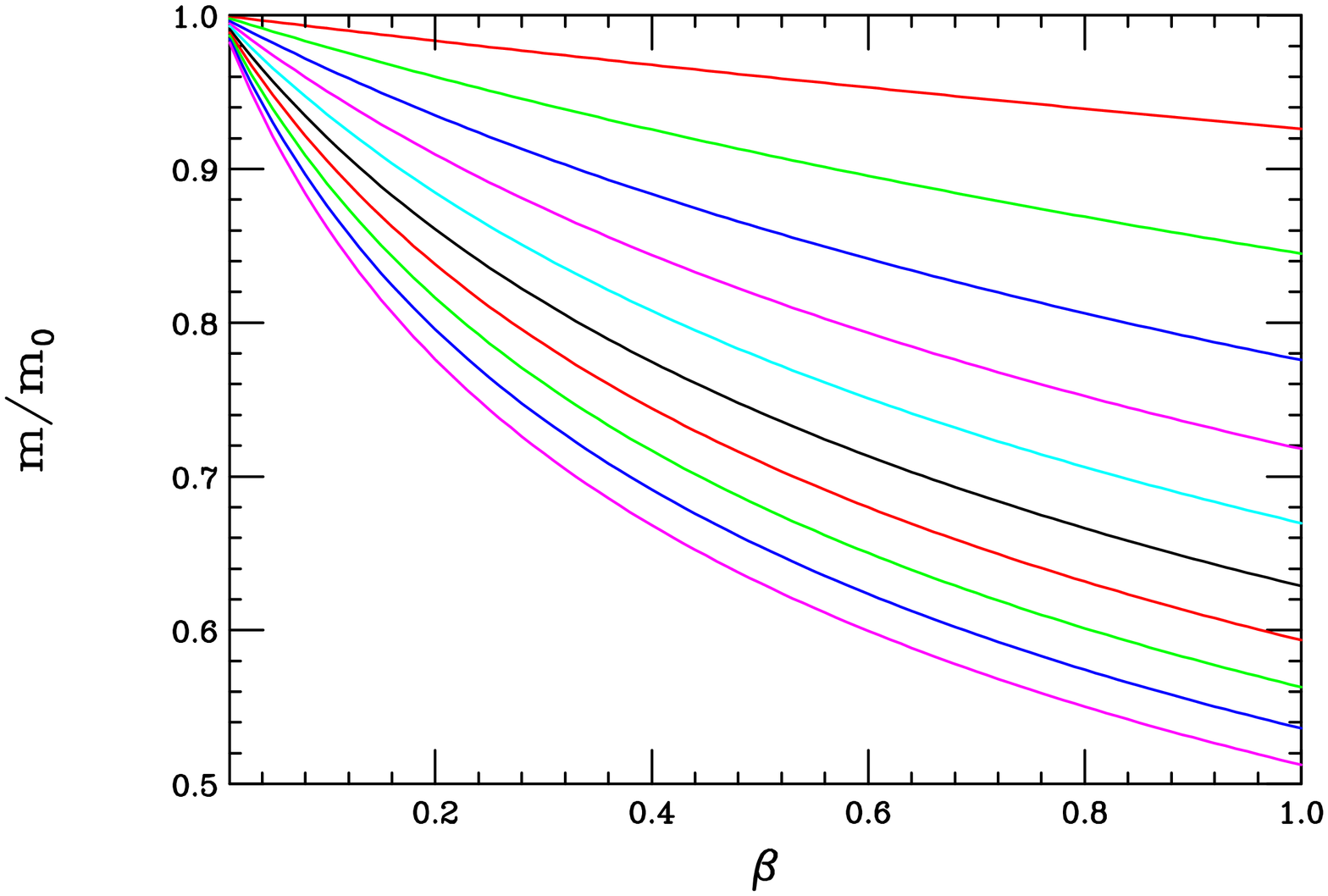,height=7.5cm,width=7.5cm,angle=90}
}
\vspace*{0.20cm}
\caption{Shifts in the RS KK graviton 
masses due to finite $\beta$ corrections: 
(Left) As a function of $c$ with $\beta=0.1$ to 1 in steps of 0.1 from top to 
bottom and (Right) as a function of $\beta$ for $c=0.01-0.1$ in steps of 0.01 
from top to bottom.}
\label{fig1}
\end{figure}

The ADD case is particularly easy to analyze as the space is flat and there is 
no bulk cosmological constant. Here the expansion of $F$ to quadratic order is 
quite simple: $F\to F_R R+\big[-F_Q+F_{RR}/2\big]R^2+F_Q G$. 
 
In ADD there are two major modifications due to $R\to F$. First, due to the 
existence of the D-dimensional bulk scalar which produces its own KK tower 
there are new exchanges between SM sources and new emissions that can lead to 
missing energy signatures. However, the effect of these new states on these 
conventional 
signatures is quite suppressed as scalars couple to the trace of the energy 
momentum tensor. For SM fields the ratio 
$T^2/T_{\mu\nu}T^{\mu\nu} \sim (m^2/s)^k$, with 
$m$ a SM particle mass and $k=1,2$. 
Further suppression occurs since the KK spectrum of the 
scalars begins at the value of the bulk mass $m_S$; for most reasonable 
choices of $F$ it can be shown that $m_S$ is naturally $\sim M \sim$ TeV. 
For example, if $F=R+\beta R^2/M^2$, we find that the value of $m_S^2$ is    
$(n+2)M^2/(4\beta (n+3))$; thus we see that 
with $\beta$ of O(1), $m_S \sim M$. 

The second effect is to modify the usual ADD relation to 
\begin{equation}
\mpl^2=V_nM^{n+2}F_R\,,
\end{equation}
where we expect 
$F_R$ is O(1) and $V=(2\pi R)^n$ for a torus. Since \mpl is fixed, for a 
given $M$ the effect of $F_R \neq 1$ is to shift the value of $R$ which leads 
to a modification of the KK spectrum as $m_{KK}\sim 1/R$, \ie, $m_{KK} 
\to m_{KK}F_R^{1/n}$. Similar arguments lead to rescalings of the graviton 
emission amplitude, $d\sigma_{ADD} \to F_R^{-1} d\sigma_{ADD}$ as well as the 
amplitude for graviton KK exchange, $A_{KK}\to F_R^{-1}A_{KK}$, both of which 
can lead to important experimental consequences, \ie, O(1) modifications to 
anticipated ADD search reaches at colliders. 
\begin{figure}[htbp]
\centerline{
\psfig{figure=shift3.ps2,height=7.5cm,width=7.5cm,angle=90}
\hspace*{5mm}
\psfig{figure=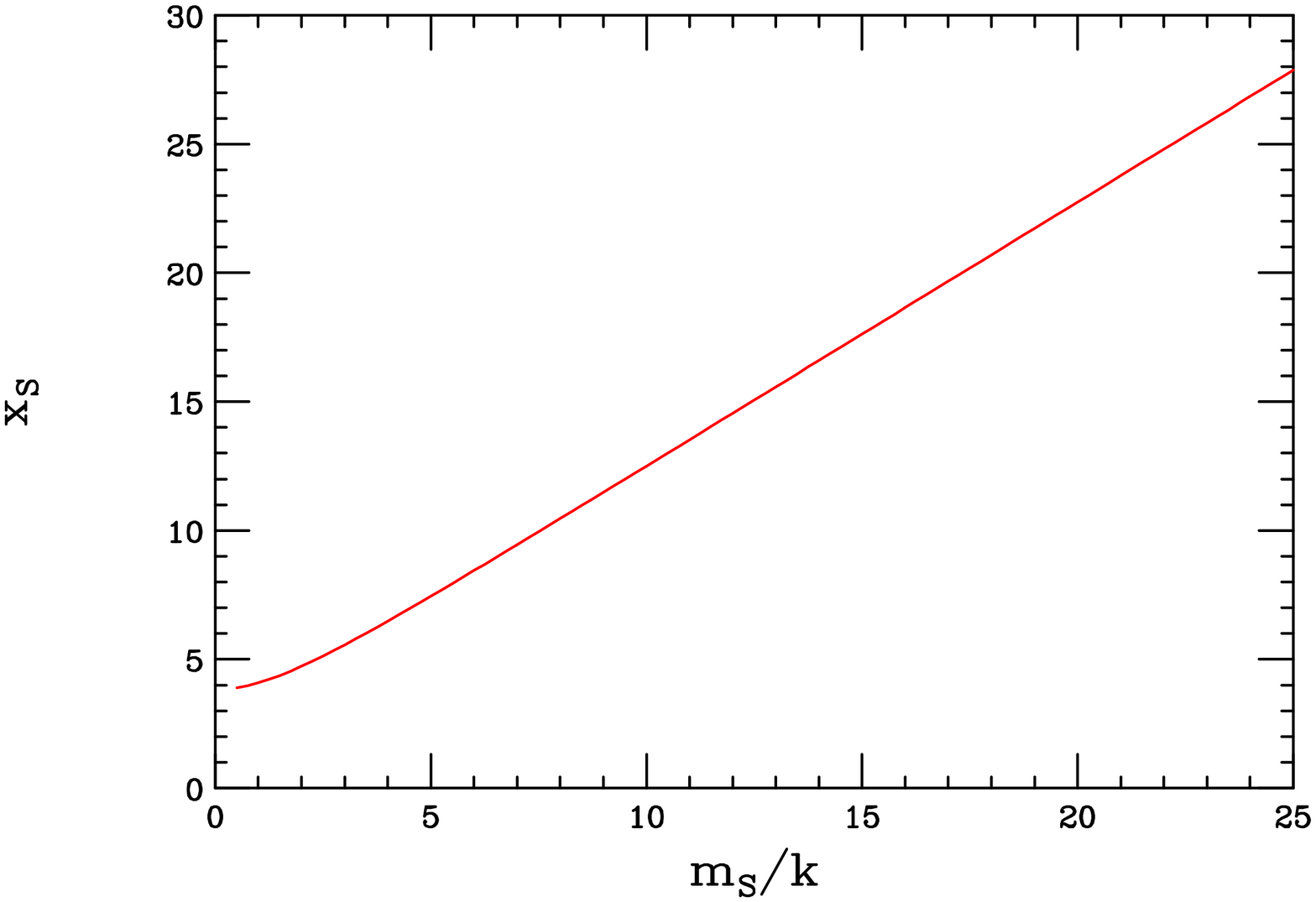,height=7.5cm,width=7.5cm,angle=90}
}
\vspace*{0.20cm}
\caption{(Left) Shift in the cosmological constant in RS due to finite 
$\beta$ for $c=0.01-0.10$ from top to bottom in 0.01 steps. (Right) 
Root for the determination of mass of the lightest KK state 
corresponding to the new RS scalar as a function of the scaled bulk mass. 
As a point of comparison the first KK graviton has a root of $\simeq$ 3.83.}
\label{fig2}
\end{figure}

\section{RS}

In RS the bulk is of constant background curvature with $R_0=-20k^2$, $k$ 
being defined via the metric $ds^2=e^{-2k|y|}\eta_{\mu\nu}dx^\mu dx^\nu-dy^2$ 
induced by a bulk cosmological constant $\Lambda$. For general $F(R,Q-4P)$, 
the equations of motion lead to a relationship between $F,k,M$ and $\Lambda$: 
$224k^4F_Q+8k^2F_R +F_0=2\Lambda/M^3$ from which $k$ as a function of 
$\Lambda$ (or the other way around) can be extracted. It is important to 
remember here that $F_{0,R,Q}$ are themselves functions of $k$. For example, 
if we assume that $F=R+R^2/M^2$, we obtain  
\begin{equation}
k^2={{3M^2}\over {40\beta}}\Big[1\pm \Big(1+{{40\Lambda}\over {9M^5}}
\beta\Big)^{1/2}\Big]\,,
\end{equation}
for which the negative root goes over to the usual EH result as the parameter 
$\beta \to 0$. Further analysis of the equations of motion shows that the 
basic RS relationship between  the 5d fundamental scale and \mpl is also 
altered by $F \neq R$; we now find that  
\begin{equation}
H(k){{M^3}\over {k}}=\mpl^2\,, 
\end{equation}
where in general $H=F_R+36k^2F_Q+1000k^4F_{RQ}+10080k^6F_{QQ}$. If, for 
example, 
$F=R+R^2/M^2$, then $H=1-40\beta k^2/M^2$. Since \mpl is known, for a fixed 
value of $M$ the two expressions above lead to shifts in the values of both 
$k$ 
and $\Lambda$ as compared to standard RS expectations and, consequently, to a 
shift in the KK graviton 
spectrum.  This is shown in Fig.~\ref{fig1} for our simple 
example of $F=R+\beta R^2/M^2$ where the KK mass shift is explicitly given by 
\begin{equation}
{m\over {m_0}}= (80\beta c^{4/3})^{-1}\big[-1+(1+160\beta c^{4/3})^{1/2}
\Big]\,.
\end{equation}
Here we have defined $c=k_0/\mpl$, where $k_0$ is the value obtained 
in the usual RS model employing the EH action. The corresponding shift in 
$\Lambda$ is shown in Fig.~\ref{fig2}. 
\begin{figure}[htbp]
\centerline{
\psfig{figure=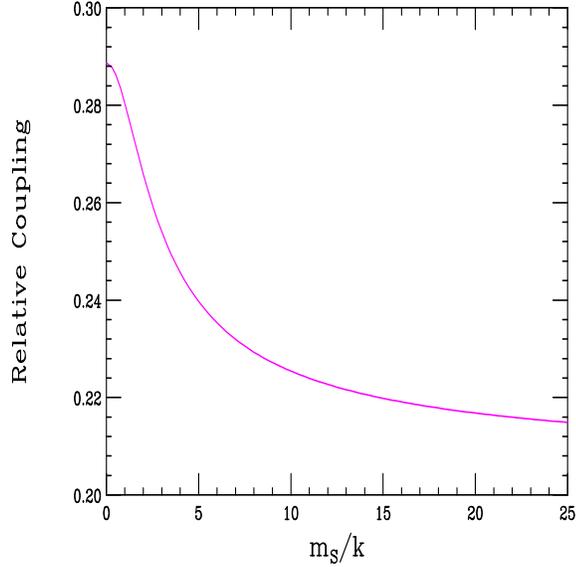,height=7.5cm,width=7.5cm,angle=90}
}
\vspace*{0.20cm}
\caption{Estimate of the relative wavefunction suppression for the lightest 
scalar KK coupling to matter in comparison to the corresponding KK graviton. 
Recall that scalars couple to $T$ whereas gravitons couple to $T_{\mu\nu}$.}
\label{fig3}
\end{figure}

As in ADD, the new scalar obtains a bulk mass which in RS case is given by  
\begin{equation}
m_S^2={3\over {8}}~{{F_R+20k^2F_{RR}+280k^4F_{RQ}}\over {F_{RR}-2F_Q-40k^2
F_{RQ}-560k^4F_{QQ}}}\,. 
\end{equation}
Within a given model this value can be used to obtain the scalar KK spectrum 
by solving for the roots of the equation 
$(2-\nu)J_\nu(x_n)+x_nJ_{\nu-1}(x_n)=0$ with $J$ the usual Bessel function and 
$\nu^2=4+m_S^2/k^2$ in the usual RS fashion; the KK masses are then given by 
$m_n=x_nke^{-\pi kr_c}$. Fig.~\ref{fig2} shows the root for the lightest KK 
scalar state, $x_S$, as a function of $m_S/k$. We thus naturally 
expect these scalar KK states to be more massive than the corresponding 
more familiar graviton excitations. Since their couplings are suppressed as 
in the ADD case they might be difficult to detect experimentally; this is 
made even more so by the small value of the scalar KK wavefunction evaluated 
at the TeV brane where the SM matter is assumed to be localized. This further  
relative suppression is shown in Fig.~\ref{fig3}. 

Hopefully new dimensions will be discovered at the Terascale.

\vskip0.2cm

\noindent{\Large\bf Acknowledgments}


This work supported in part by the Department of Energy, Contract 
DE-AC02-76SF00515.

\end{document}